\documentclass[%
superscriptaddress,
showpacs,
amsmath,amssymb,
twocolumn,
prc
floatfix,]
{revtex4-1}
\usepackage{color}
\definecolor{mygray}{gray}{.9}
\definecolor{mypink}{rgb}{.99,.91,.95}
\definecolor{mycyan}{cmyk}{.3,0,0,0}
\usepackage{graphicx}
\usepackage{dcolumn}
\usepackage{bm}
\usepackage{amsmath,amssymb}
\usepackage{float}
\usepackage{booktabs}
\usepackage{natbib}

\newcommand{\element}[2]{~${}^{#1}${#2}}

\begin{document}
	
	
	
	\title{ Parity-doublet bands in the odd-\bm{$A$} isotones \element{237}U and \element{239}Pu by a particle-number-conserving method based on the cranked shell model}
	\author{Jun Zhang}
	\affiliation{College of Science, Nanjing University of Aeronautics and Astronautics, Nanjing 210016, China}
	\author{Xiao-Tao He}
	\email{hext@nuaa.edu.cn}
	\affiliation{College of Materials Science and Technology, Nanjing University of Aeronautics and Astronautics, Nanjing 210016, China}

	\date{\today}
\begin{abstract}
Based on the reflection-asymmetric Nilsson potential, the parity-doublet rotational bands in odd-$A$ isotones \element{237}U and \element{239}Pu have been investigated by using the particle-number-conserving (PNC) method in the framework of the cranked shell model (CSM). The experimental kinematic moments of inertia (MOIs) and angular momentum alignments are reproduced very well by the PNC-CSM calculations. The significant differences of rotational properties between \element{237}U and \element{239}Pu are explained with the contribution of nucleons occupying proton octupole-correlation pairs of $\pi ^{2} i_{13/2}f_{7/2}$. The upbendings of moments of inertia of the parity-doublet bands in \element{237}U are due to the interference terms of alignments of protons occupying $\pi f_{7/2}$ (${1/2}$) and the high-$j$ intruder $\pi i_{13/2}$ $(1/2, 3/2)$ orbitals. The splittings between the simplex partner bands of the parity-doublet bands in both \element{237}U and \element{239}Pu result from the contribution of alignment of neutron occupying the $\nu d_{5/2}$ ($1/2$) orbital.
\end{abstract}
	\maketitle
\section{Introduction}
The typical feature for octupole correlations in odd-mass nuclei is the appearance of parity-doublet rotational bands, which are pairs of almost degenerate states in excitation energy with the same spin but opposite parities \cite{SahaS1980_PRC21_2322,SoodP1989_PRC40_1530,GrafenV1991_PRC44_1728,ReviolW2009_PRC80_11304}. 
The octupole correlations in nuclei are associated with the single-particle states with orbital and total angular momentum differing by 3, i.e., $\Delta{l} = \Delta{j} = 3$ \cite{AhmadI1993_ARoNaPS43_71,AbergS1990_ARoNaPS40_439}. The nuclei with $Z \approx 88$ and $N \approx 134$ in the actinide region are expected to possess the maximum octupole correlations since their Fermi surface lies between the proton $\pi ^{2}f_{7/2}i_{13/2}$ and neutron $\nu ^{2}g_{9/2}j_{15/2}$ octupole correlation pairs \cite{ButlerP1996_RoMP68_349}. 
Experimentally, since the first observation of parity-doublet rotational bands in odd-mass \element{229}Pa and \element{227}Ac \cite{AhmadI1982_PRL49_1758,Martz.H_1988_PRC37_1407},  similar bands have been observed in many odd-mass actinide nuclei, like \element{219,223,225}Ac, \element{219,221,223}Fr, \element{221,223}Th, and so on \cite{DrigertM1985_PRC31_1977,Parr.E_2022_PRC105_34303,Ahmad.I_1984_PRL52_503,Liang.C_1991_PRC44_676,Ardisson.G_2000_PRC62_64306,
Sheline.R_1995_PRC51_1708,Reviol.W_2014_PRC90_44318,Dahlinger.M_1988_NPA484_337}.\par
The parity-doublet bands are observed experimentally in \element{237}U and \element{239}Pu, in which some interesting properties are reported \cite{Wiedenhoever.I_1999_PRL83_2143,ZhuS2005_PLB618_51}.   In Ref. \cite{Wiedenhoever.I_1999_PRL83_2143}, the angular momentum alignments were compared for the bands in the Pu isotopes with $238 \leqslant A \leqslant 244$, where the sharp backbending observed in the heavier isotopes is not present within the same frequency range in \element{239}Pu. The suddenly gained alignments in the heavier isotopes are due to the contribution from a pair of $i_{13/2}$ protons. There is at present no satisfactory explanation for the absence of backbending phenomenon in \element{239}Pu. As the $N = 145$ isotones closest to the \element{239}Pu, the behavior of the alignments with rotational frequency  for \element{237}U is very different. In Ref. \citep{ZhuS2005_PLB618_51}, a strong backbending in alignment occurs at $\hbar\omega \approx$ 0.25 MeV in \element{237}U. In both work, the experimental observations support the presence of the large octupole correlations in the ground states and in the low-lying excited states of odd-$A$ \element{237}U and \element{239}Pu nuclei. The spin and parity are assigned as $K^{\pi} = 1/2^{+}$ for the ground-state bands in both nuclei. The experimental data shows that there exist significant simplex splittings at low rotational frequency region in the parity-doublet bands in both nuclei. These issues need further investigations.\par 
Many theoretical approaches were developed to investigate the octupole correlations in nuclei. These include the reflection-asymmetric mean-field approach \cite{Bender.M_2004_PRC70_54304,Robledo.L_2011_PRC84_54302,Zhang.W_2010_PRC81_34302}, algebraic models \cite{Ganev.H_2004_PRC69_14305,Solnyshkin.A_2005_PRC72_64321}, cluster models \cite{Chen.X_2016_PRC94_21301,Delion.D_2012_PRC85_64306,Warda.M_2011_PRC84_44608}, vibrational approaches \cite{Elvers.M_2011_PRC84_54323,Dobrowolski.A_2018_PRC97_24321,Bizzeti.P_2010_PRC81_34320}, reflection asymmetric shell model \citep{Chen.Y_2000_PRC63_14314,
Gao.Z_2011_PRC83_57303,Gao.Z_2006_PRC74_54303} and cranked shell model \cite{HuebelH1986_NPA453_316,BengtssonR1986_ADaNDT35_15,Heenen.P_1994_PRC50_802,NazarewiczW1985_NPA441_420,HeX2020_PRC102_64328}. The particle-number-conserving method in the framework of cranked shell model (PNC-CSM) is one of the most useful models to describe the rotational bands \cite{ZengJ1994_PRC50_1388,ZhangZ2009_PRC80_34313,HeX2005_NPA760_263}. In the PNC-CSM calculations, the cranked shell model Hamiltonian is diagonalized directly in a truncated Fock space. The particle number is conserved exactly and the Pauli blocking effects are taken into account spontaneously. The previous PNC-CSM method has been developed to deal with reflection-asymmetric nuclei, and successfully applied to describe the alternating-parity rotational bands in the even-even nuclei \cite{HeX2020_PRC102_64328}.\par  
In this work, the PNC-CSM method is further applied to describe the ground-state parity-doublet rotational bands in odd-$A$ nuclei \element{237}U and \element{239}Pu. The ground-state spin assignments are identified for both nuclei. By considering the octupole correlations, Pauli blocking effects and Coriolis interaction, the striking differeces of rotational properties between \element{237}U and \element{239}Pu and the simplex splittings of the parity-doublet bands in both nuclei are investigated in detail.\par
A brief introduction of the PNC-CSM method dealing with the reflection-asymmetric nuclei are presented in Sec. \ref{Sec:PNC}. The detailed PNC-CSM analyses for the parity-doublet rotational bands in \element{237}U and \element{239}Pu are presented in Sec. \ref{Sec:results}. A summary is given in Sec. \ref{Sec:summary}. \par
\section{Theoretical framework}{\label{Sec:PNC}}
The detailed understanding of PNC-CSM method dealing with the reflection-asymmetric nucleus can be found in Ref. \cite{HeX2020_PRC102_64328}. Here we give a brief description of the related formalism. The cranked shell model Hamiltonian of an axially symmetric nucleus in the rotating frame \cite{ZengJ1983_NPA405_1,LiuS2002_PRCNP66_024320,ZhangZ2016_NPA949_22} is
\begin{equation}
	 H_\mathrm{CSM} = H_\mathrm{0} + H_\mathrm{P} = H_{\rm Nil} -\omega J_x + H_\mathrm{P}.  
\end{equation}
$H_{\rm Nil} = \sum h_{\rm Nil}(\varepsilon_2, \varepsilon_3, \varepsilon_4)$ is the Nilsson Hamiltonian, where quadrupole ($\varepsilon_2$), octupole ($\varepsilon_3$), and hexadecapole ($\varepsilon_4$) deformation parameters are included. $-\omega J_x = -\omega \sum j_{x}$ repesents the Coriolis interaction with the rotational frequency $\omega$ about the $x$ axis (perpendicular to the nucleus symmetry $z$ axis). \par
When $\hbar\omega = 0$, for an axially symmetric and reflection-asymmetric system, the single-particle Hamiltonian has nonzero octupole matrix elements between different shell $N$. Since parity $p = (-1)^{N}$, the parity is no longer a good quantum number, but the single-particle angular momentum projection on the symmetry axis $\Omega$ is still a good quantum number. The single-particle orbitals can be labelled with the quantum numbers $\Omega$ ($l_j$), where $l_j$ are the corresponding spherical quantum numbers.
\par 
However, when $\hbar\omega \neq {0}$, due to the Coriolis interaction $-\omega j_{x}$, the $\Omega$ is no longer a good quantum. Since the reflection through plane $yoz$, $S_x$ invariant holds \cite{HornyakW2012_unknown_}. The single-particle orbitals can be labelled with the simplex quantum numbers $s$ ($s =\pm i$), which are the eigenvalues of $S_x$ operator.
\par
The pairing $H_\mathrm{P}$ includes the monopole and quadrupole pairing correlations $H_\mathrm{P}(0)$ and $H_\mathrm{P}(2)$,
\begin{equation}
	H_\mathrm{P}(0) = -G_{0} \sum_{\xi\eta}a_{\xi}^{\dagger}a_{\overline{\xi}}^{\dagger}a_{\overline{\eta}}a_{\eta} ,
\end{equation}
\begin{equation}
	H_\mathrm{P}(2) = -G_{2} \sum_{\xi\eta} q_{2}(\xi)q_{2}(\eta)a_{\xi}^{\dagger}a_{\overline{\xi}}^{\dagger}a_{\overline{\eta}}a_{\eta} ,	
\end{equation}
where $\overline{\xi} (\overline{\eta})$ labels the time-reversed state of a Nilsson state $\xi(\eta)$, $q_{2}(\xi) = \sqrt{{16\pi}/{5}} \langle \xi |r^{2}Y_{20}|\xi\rangle$ is the diagonal element of the stretched quadrupole operator, and $G_{0}$ and $G_{2}$ are the effective strengths of monopole and quadrupole pairing interactions, respectively.\par
By diagonalizing $H_\mathrm{CSM}$ in a sufficiently large cranked many-particle configuration (CMPC) space, a sufficiently accurate low-lying excited eigenstate is obtained as
\begin{equation}
	|\psi\rangle = \sum_{i}C_{i}|i\rangle ,
\end{equation}
where $C_{i}$ is real and $|i\rangle = |\mu_{1}\mu_{2}\dots\mu_{n}\rangle$ is a cranked many-particle configuration for an $n$-particle system, and $\mu_{1}\mu_{2}\dots\mu_{n}$ are the occupied cranked Nilsson orbitals. The configuration $|i\rangle$ is characterized by the simplex $s_{i}$,
\begin{equation}
	s_{i}=s_{\mu_{1}} s_{\mu_{2}}  \dots s_{\mu_{n}}
\end{equation}
where $s_{\mu}$ is the simplex of the particle occupying in orbital $\mu$.\par
The occupation probability of each cranked orbital $\mu$ can be calculated as:    \par
\begin{equation}
	n_{\mu} = \sum_{i}{|C_{i}|}^{2}P_{i\mu} ,
\end{equation}
here $P_{i\mu} = 1$ if $\mu$ is occupied and $P_{i\mu} = 0$ otherwise.\par 
The angular momentum alignment of eigenstate, including the diagonal and off-diagonal parts, can be written as
\begin{equation}
	\langle\psi|J_x|\psi\rangle = \sum_{i}{|C_{i}|}^{2}\langle i|J_x|i \rangle + \sum_{i \neq j}{C_{i}}^{*}{C_{j}}\langle i|J_x|j \rangle ,
\end{equation}
and the kinematic moment of inertia is
\begin{equation}
	J^{(1)} = \dfrac{1}{\omega}\langle\psi|J_x|\psi\rangle .
\end{equation}  \par
For reflection-asymmetric systems with odd number of nucleons \cite{HornyakW2012_unknown_}, the experimental rotational band with simplex $s$ is characterized by spin state $I$ of alternating parity,
\begin{eqnarray}
	&&s=+i,I^{p}=1/2^{+},3/2^{-},5/2^{+},7/2^{-},\cdots,\\
	&&s=-i,I^{p}=1/2^{-},3/2^{+},5/2^{-},7/2^{+},\cdots,
\end{eqnarray}  \par
The angular momentum alignment for positive- and negative-parity bands can be expressed as \cite{R.V.JolosP1995_NPA587_377}
\begin{equation}
	\langle J_x\rangle_{p}=\langle\psi|J_x|\psi\rangle-\frac{1}{2}p\triangle I_{x}(\omega),
\end{equation}
\begin{equation}
	J^{(1)}_{p}=\frac{\langle\psi|J_x|\psi\rangle}{\omega}-\frac{1}{2}p\triangle J^{(1)}(\omega),
\end{equation}
where $|\psi\rangle$ is the parity-independent wave function with the rotational frequency $\omega$ calculated by PNC-CSM method. The $\triangle I_{x}(\omega)$ and $\triangle J^{(1)}(\omega)$ are parity splitting of the alignment and moment of inertia in the experiment rotational bands, respectively. The corresponding values can be obtained by the following formula,
\begin{equation}
	\triangle I_{x}(\omega)=I_{x-}(\omega)-I_{x+}(\omega),
\label{eq:deltaI}
\end{equation}
\begin{equation}
	\triangle J^{(1)}(\omega)=J^{(1)}_{-}(\omega)-J^{(1)}_{+}(\omega),
\label{eq:deltaJ}
\end{equation}
in which $+(-)$ represents the positive- (negative-) parity in the experiment rotational bands.

\section{Results and discussions}{\label{Sec:results}}
\subsection{Parameters}
In the present calculation, the values of Nilsson parameters ($\kappa,\mu$) are taken from Ref. \cite{ZhangZ2012_PRC85_14324}. The deformation parameters $\varepsilon_2$, $\varepsilon_3$, $\varepsilon_4$ are input parameters in the PNC-CSM calculations, the values employed in this work are listed in Table \ref{deformation parameters}. The deformation parameters $\varepsilon_2$ and $\varepsilon_4$ for nuclei \element{237}U and \element{239}Pu are taken as an average of the neighboring even-even U (Z=92) and Pu (Z=94) isotopes \cite{HeX2020_PRC102_64328}. The $\varepsilon_3$ values used in this work are chosen by fitting the experimental MOIs and alignments of the ground-state bands in \element{237}U and \element{239}Pu. \par
The effective pairing strengths $G_{0}$ and $G_{2}$, in principle, can be determined by the odd-even differences in nuclear binding energies, and are connected with the dimensions of the truncated CMPC space. For both \element{237}U and \element{239}Pu, the CMPC space is constructed in the proton $N=5$, 6 shells and the neutron $N=6$, 7 shells. The dimensions of the CMPC space are about 1000 for both protons and neutrons. The corresponding effective monopole and quadrupole pairing strengths are $G_{0p}$   = 0.25 MeV and $G_{2p}$ = 0.03 MeV for protons, $G_{0n}$ = 0.15 MeV and $G_{2n}$ = 0.01 MeV for neutrons. In this work, the same values of effective pairing parameters are used for \element{237}U and \element{239}Pu. The stability of the PNC-CSM calculations against the change of the dimensions of the CMPC space has been investigated in Ref. \cite{ZengJ1994_PRC50_1388}.\par
\begin{table}[!htbp]
	\caption{Deformation parameters $\varepsilon_2$, $\varepsilon_3$, $\varepsilon_4$ used in the present PNC-CSM calculations for \element{237}U and \element{239}Pu}
	\setlength{\tabcolsep}{6.5mm}
	\renewcommand{\arraystretch}{1.25}
	{
		\begin{tabular}{l c c c }
			\hline\hline
			
			& $\varepsilon_2$ & $\varepsilon_3$  & $\varepsilon_4$  \\[3pt]
			\hline
			
			\element{237}U    &0.210       &0.071        &-0.045\\[3pt]
			\element{239}Pu   &0.230       &0.010        &-0.055\\[3pt]
			
			\hline\hline
		\end{tabular}
	}
	\label{deformation parameters}
\end{table}
\begin{figure}[!htbp]
	\includegraphics[scale=0.33]{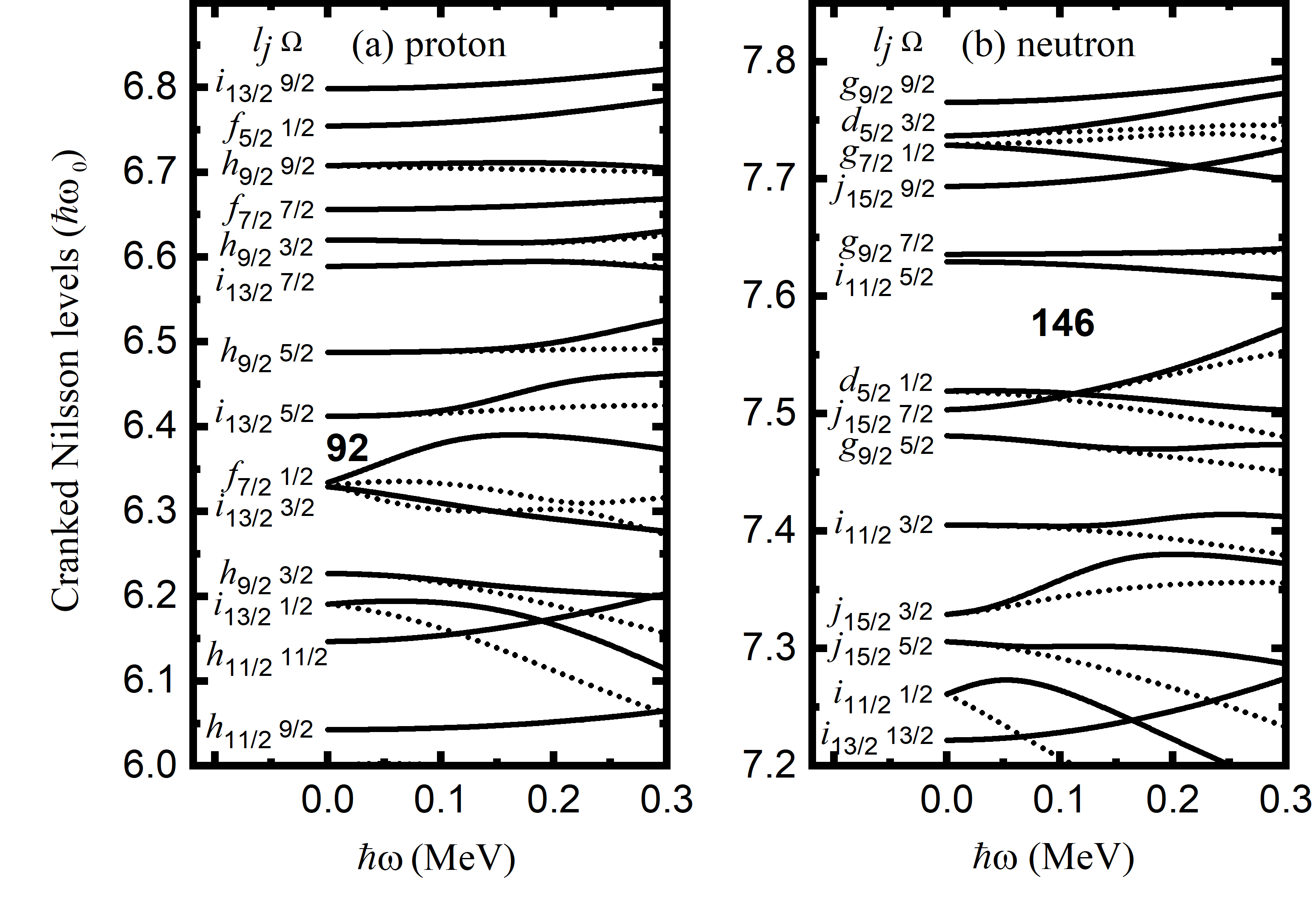}
	\caption{The cranked Nilsson levels near the Fermi surface of  \element{237}U for protons (a) and neutrons (b). The simplex $s=+i$ ($s=-i$) levels are denoted by black solid (dashed) lines.}
	\label{Fig1}
\end{figure}
\begin{figure}[!htbp]
	\includegraphics[scale=0.33]{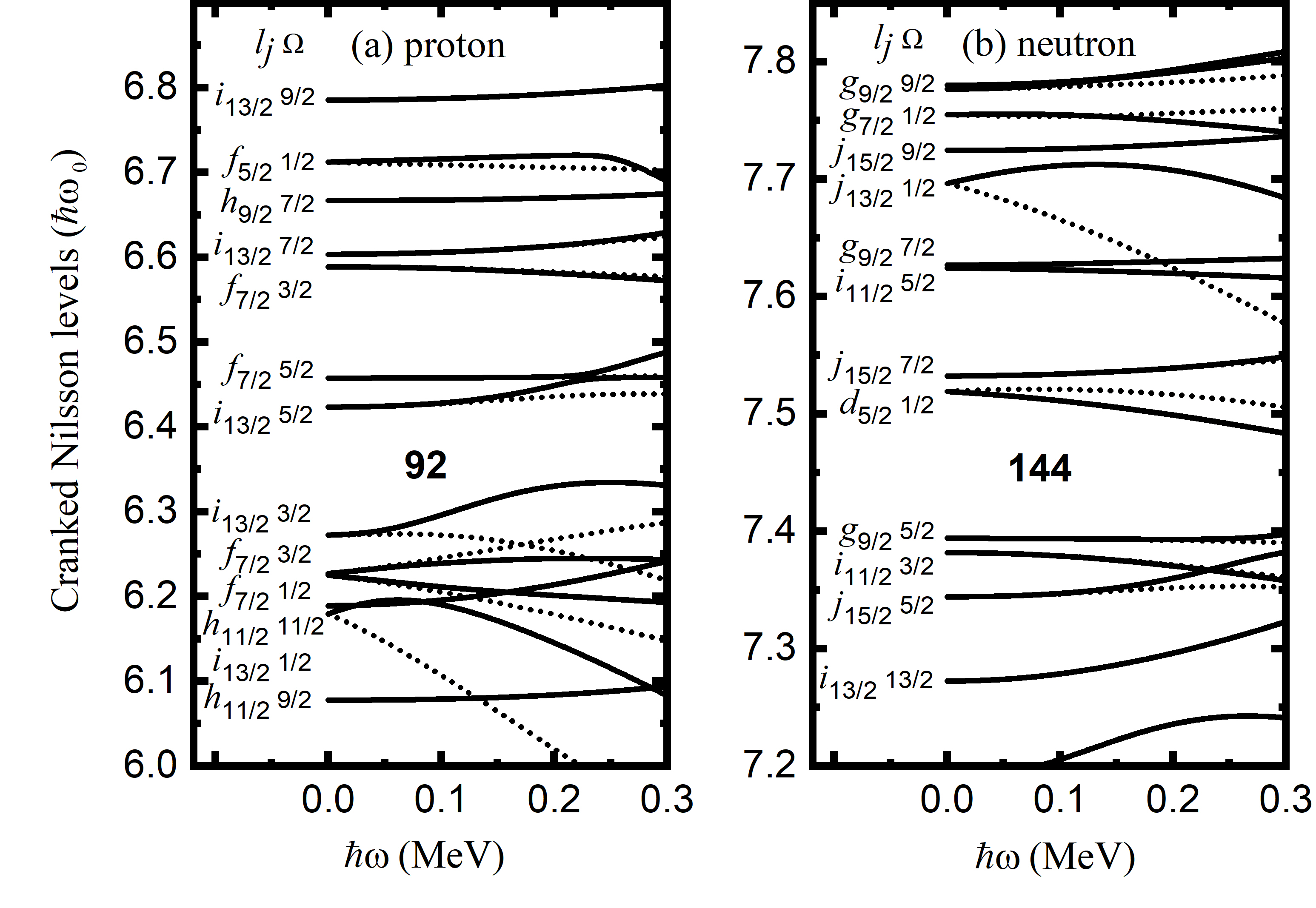}
	\caption{The same as Fig. \ref{Fig1}, but for cranked Nilsson levels near the Fermi surface of \element{239}Pu.}
	\label{Fig2}
\end{figure}

\subsection{Cranked Nilsson levels}
Figure \ref{Fig1} and \ref{Fig2} show the cranked Nilsson levels near the Fermi surface of \element{237}U and \element{239}Pu, respectively. Due to the octupole correlations, the single-particle orbitals are parity-mixed, the only conserved quantum number are $\Omega$, which represent the projection of the single-particle total angular momentum $j$ on the symmetry axis. So the parity-mixed proton and neutron orbitals are labeled by quantum number $\Omega$ at the bandhead ($\omega = 0$), and the simplex $s=+i$ ($s=-i$) levels are denoted by solid (dotted) lines. For both protons and neutrons, such a sequence of single-particle levels are found to be reasonable in agreement with experimental data taken from Refs. \citep{BasuniaM2006_NDS107_2323,BrowneE2014_NDS122_293}. The $Z = 92, 96$ gaps for protons and $N = 150, 152$ gaps for neutrons are presented in nearly all odd-mass actinides, remarkably similar to the results predicted by the Woods-Saxon potential.

Comparing the proton single-particle levels for $^{237}$U and $^{239}$Pu, the proton $\pi$1/2 level from $\pi f_{7/2}$ orbital just locates at the Fermi surface for \element{237}U, while it is far below the Fermi surface for \element{239}Pu. As we known, $\pi f_{7/2}$ and $\pi i_{13/2}$ proton orbitals form the basis for the existence of strong octupole correlations in the actinide region. The neutron $\nu$1/2 ($d_{5/2}$) orbital locates just at the Fermi surface for both \element{237}U and \element{239}Pu nuclei, which are consistent with the experimental ground state configuration with $K = 1/2$.\par
\subsection{Parity-doublet bands in \element{237}U and \element{239}Pu}

\begin{figure*}[!htbp]
	\includegraphics[scale=0.60]{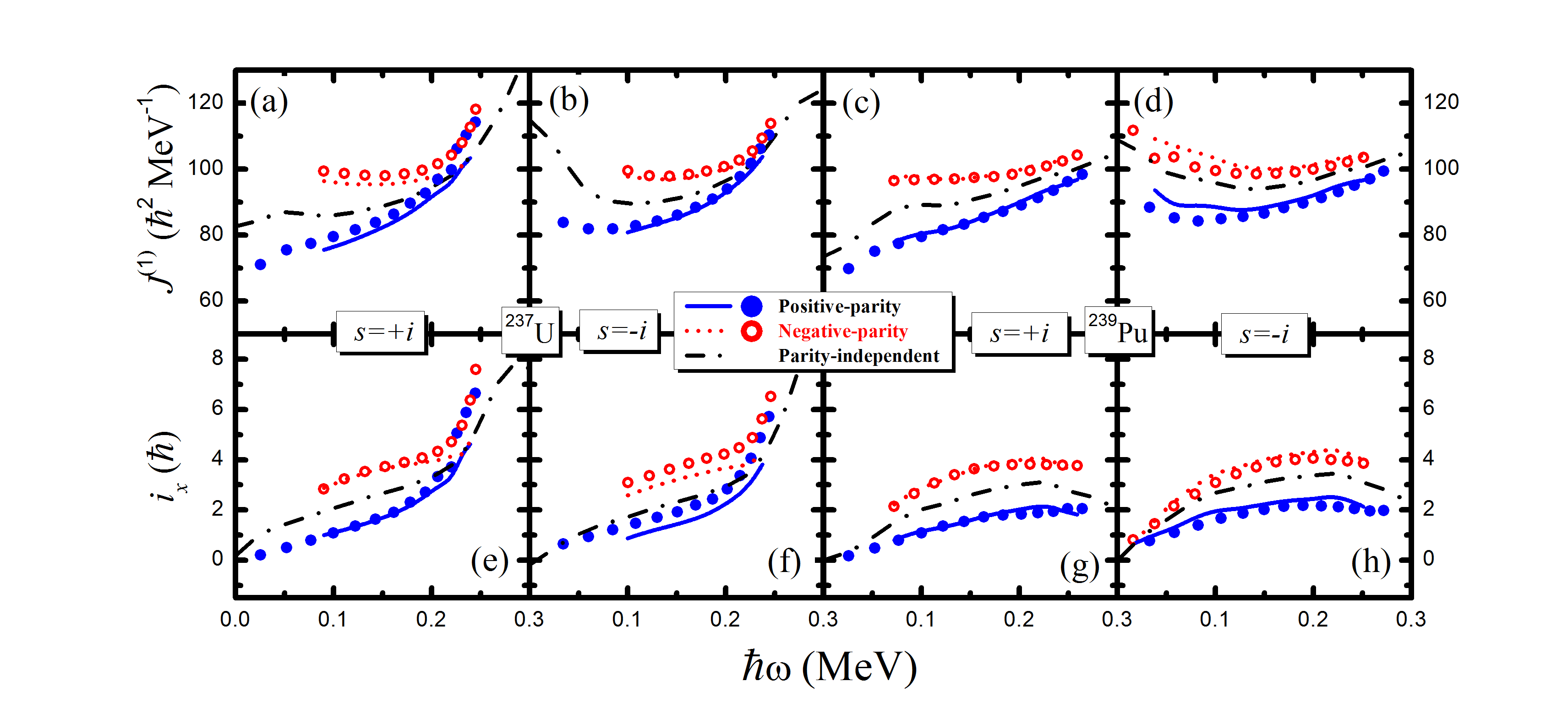}
	\caption{(Color online) The kinematic moments of inertia $J^{(1)}$ (top row) and alignments $J_x$ (bottom row) of the parity-doublet rotational bands in \element{237}U and \element{239}Pu. The experimental data are denoted by blue solid and red open circles for the positive- and negative-parity bands, respectively. The PNC-CSM calculations of parity-independent bands are denoted by dash-dotted lines. After considering the parity splitting of Eqs.~(\ref{eq:deltaJ}) and (\ref{eq:deltaI}), the positive- and negative-parity bands are denoted by blue solid and red dotted lines, respectively.}
	\label{MoI and Jx}
\end{figure*}

Figure \ref{MoI and Jx} shows the experimental and calculated MOIs and alignments of the parity-doublet rotational bands in \element{237}U and \element{239}Pu. The experimental data are taken from Refs. \cite{BasuniaM2006_NDS107_2323,BrowneE2014_NDS122_293}, which are reproduced very well by the PNC-CSM calculations. One can see that the rotational behaviors of the parity-doublet bands in \element{237}U and \element{239}Pu are very different. From Figs. \ref{MoI and Jx}(e) and \ref{MoI and Jx}(f), there are sharp upbendings at $\hbar\omega \approx$ 0.25 MeV for both the $s=+i$ and $s=-i$ alternating-parity bands in \element{237}U. In contrast, as shown in Figs. \ref{MoI and Jx}(g) and \ref{MoI and Jx}(h), the rotational behaviors are much plainer for bands in \element{239}Pu. It is well known that the backbending is caused by the crossing of the ground-state band with a pair-broken band based on high-$j$ intruder orbitals. In this mass region, the high-$j$ intruder orbitals involved in the Fermi surface are the ${\pi}i_{13/2}$ and ${\nu}j_{15/2}$ orbitals.

\begin{figure}[!htbp]
	\includegraphics[scale=0.33]{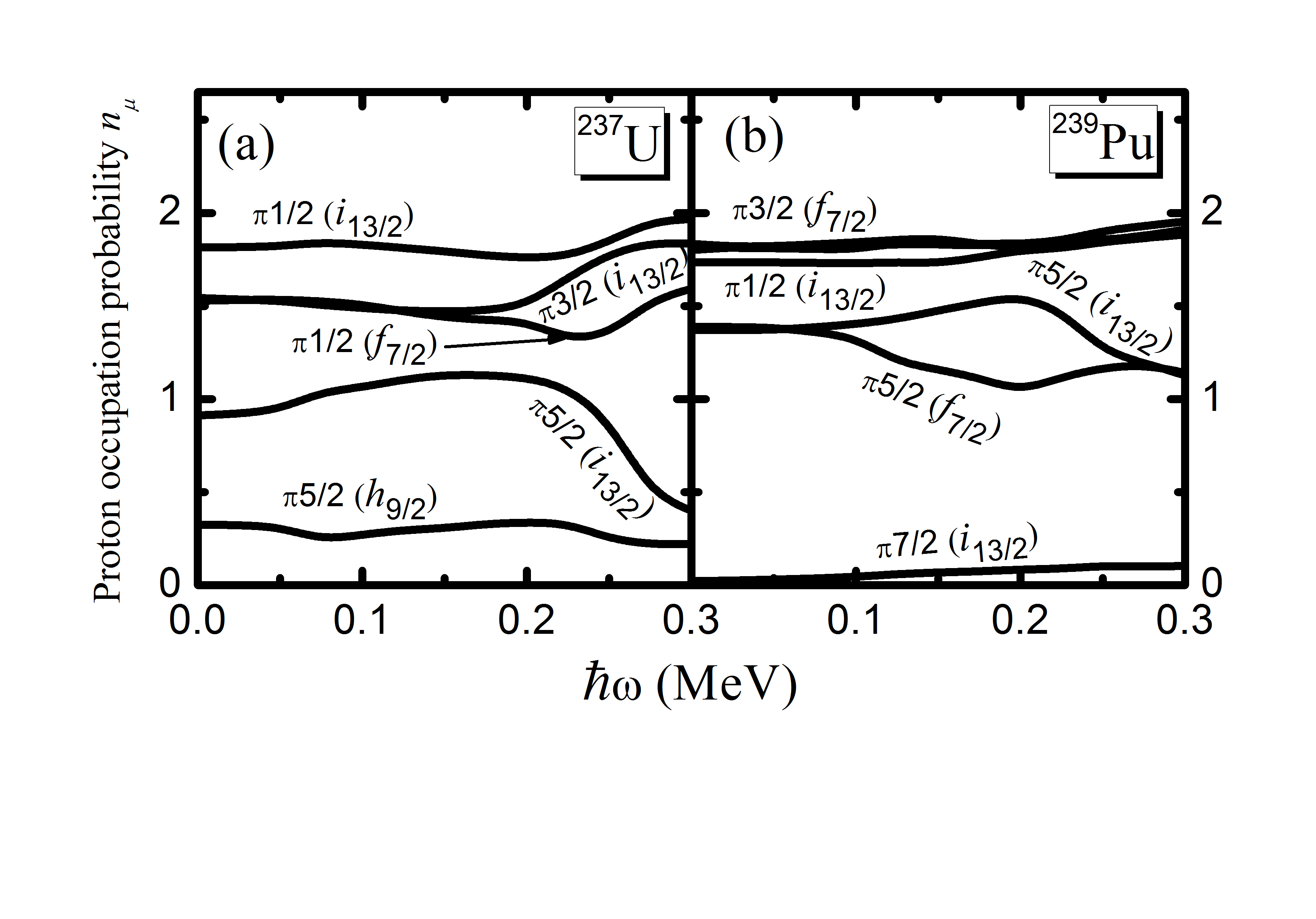}
	\caption{The occupation probability $n_{\mu}$ of each orbital ${\mu}$ (include both $s=\pm i$) for protons near the Fermi surface for the parity-doublet bands in \element{237}U and \element{239}Pu. The Nilsson levels far above ($n_{\mu} {\approx} 0$) and far below ($n_{\mu} {\approx} 2$) the Fermi surface are not shown.}
	\label{Proton_Occupation}
\end{figure}

\begin{figure}[!htbp]
	\includegraphics[scale=0.33]{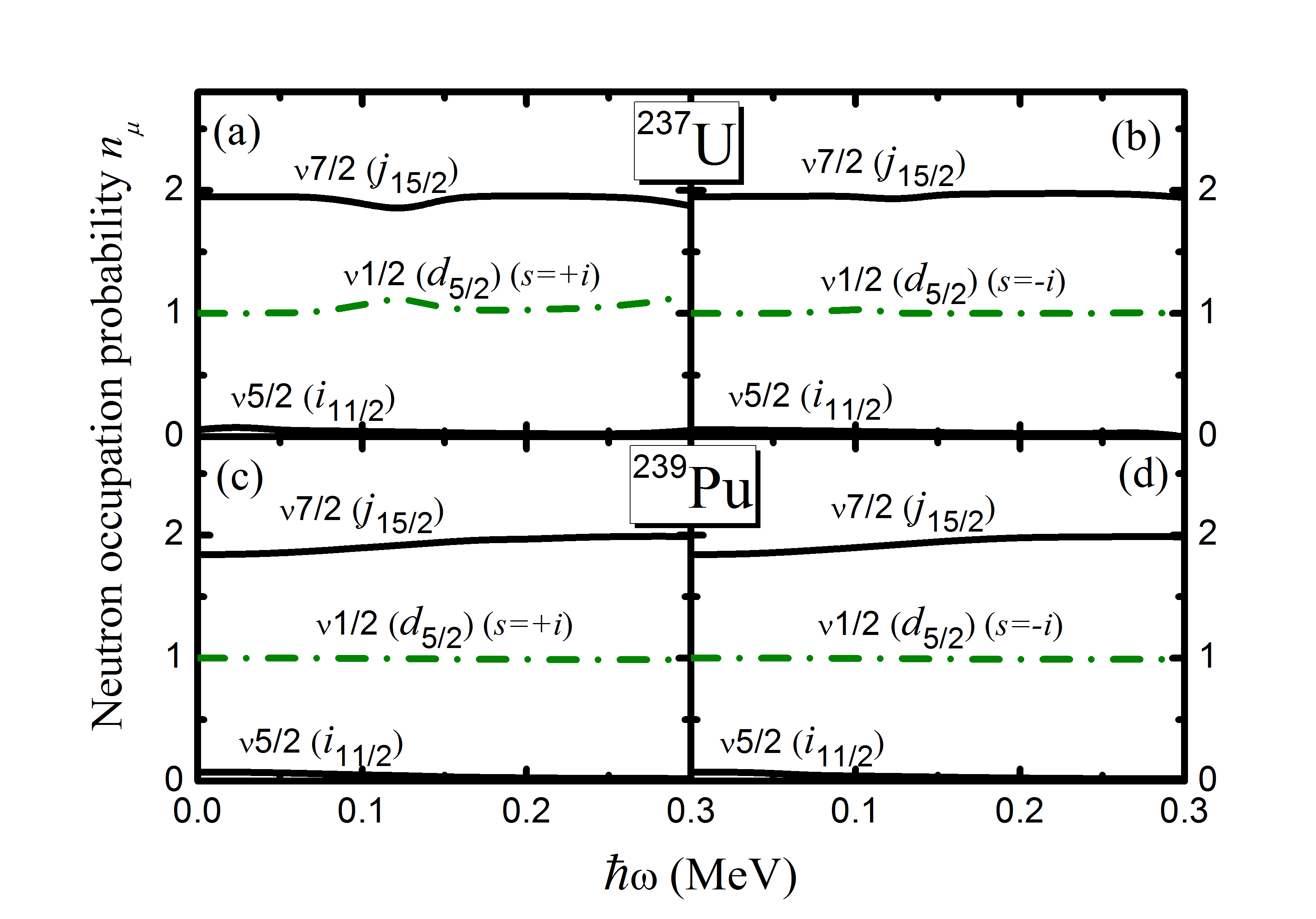}
	\caption{(Color online) The occupation probability $n_{\mu}$ of each orbital ${\mu}$ (include both $s=\pm i$ unless otherwise specified) for neutrons near the Fermi surface for the parity-doublet bands in \element{237}U (top row) and \element{239}Pu (bottom row). The Nilsson levels far above ($n_{\mu} {\approx} 0$) and far below ($n_{\mu} {\approx} 2$) the Fermi surface are not shown.}
	\label{Neutron_Occupation}
\end{figure}

Figures \ref{Proton_Occupation} and \ref{Neutron_Occupation} show the occupation probability $ n_{\mu} $ of each orbital $\mu$ (including both $s = {\pm} i$) near the Fermi surface for the parity-doublet bands in \element{237}U and \element{239}Pu for protons and neutrons, respectively. The Nilsson levels far above ($n_{\mu} {\approx}$ 0) and far below ($n_{\mu} {\approx}$ 2) the Fermi surface are not shown. In Fig. \ref{Neutron_Occupation}, the occupation probabilities of neutron orbitals in both $s=+i$ and $s=-i$ bands are almost constant with rotational frequency $\hbar\omega$ increasing, which contributes little to the $\omega$ variation of $J^{(1)}$. The different behavior of the rotational bands in \element{237}U and \element{239}Pu would be the result of the contribution from protons. 

\begin{figure}[!htbp]
	\includegraphics[scale=0.33]{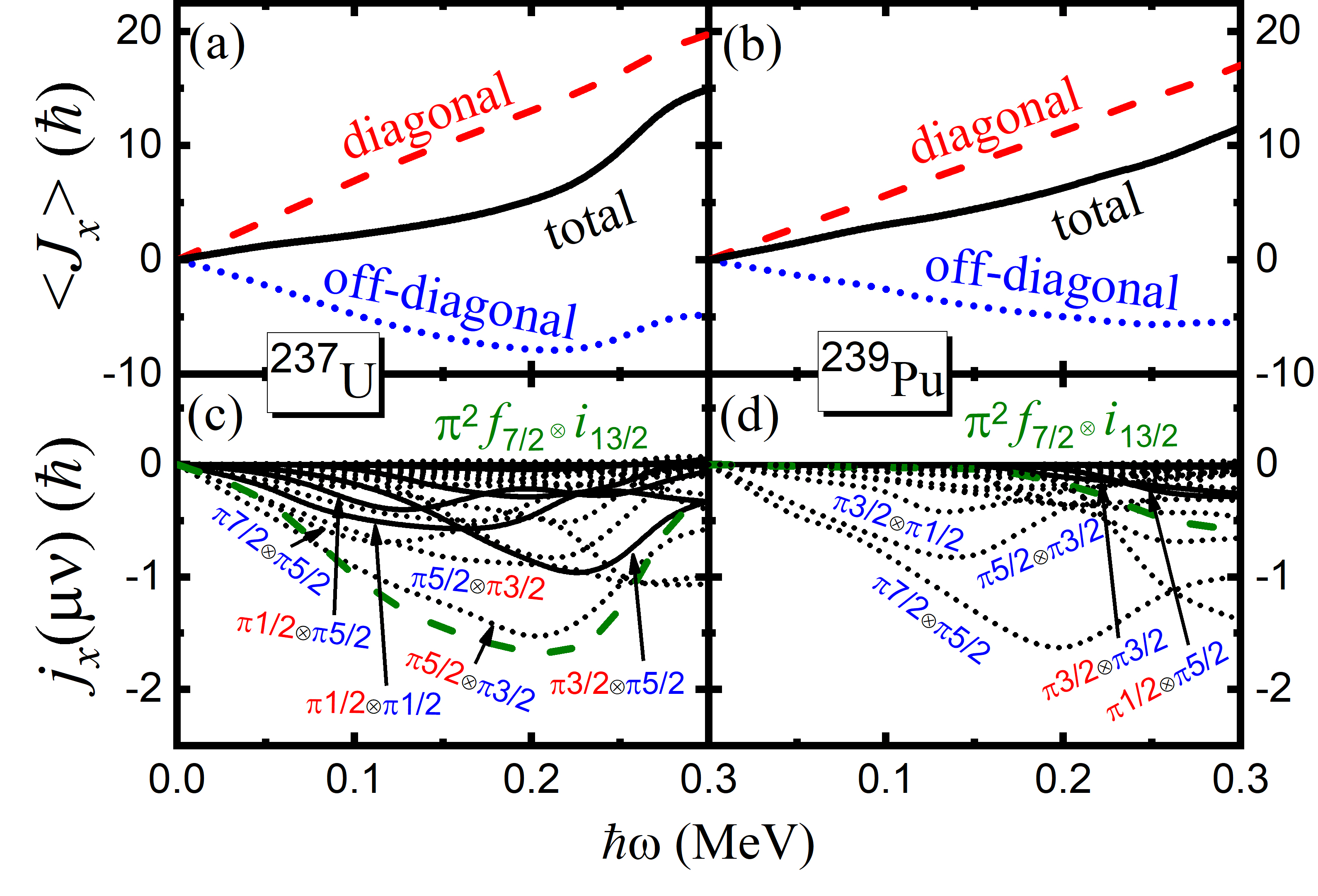}
	\caption{(Color online) Contributions of protons to the angular momentum alignments $\langle J_x \rangle$ for the parity-doublet bands in \element{237}U and \element{239}Pu. The diagonal $\sum\nolimits_{\mu}{j_x}({\mu})$ and off-diagonal parts $\sum\nolimits_{{\mu}<{\nu}}{j_x}({\mu}{\nu})$ are denoted by red dashed and blue dotted lines, respectively (top row). The interference term  $j_x(\mu\nu)$ between orbitals from proton $\pi^{2}i_{13/2}f_{7/2}$ pairs are denoted by black solid lines, other interference terms are donoted by black dotted lines. The total interference terms from proton $\pi^{2}i_{13/2}f_{7/2}$ pairs are donoted by olive dashed lines. Orbitals from $\pi f_{7/2}$ and $\pi i_{13/2}$ are denoted by red and blue quantum numbers $\Omega$, respectively (bottom row).}
	\label{Fig6}
\end{figure}

From Fig. \ref{Proton_Occupation}(a), it can be seen that the occupation probabilities of proton orbitals $\pi{1/2}$ ($f_{7/2}$), $\pi{3/2}$ ($i_{13/2}$) are almost constant ($n_{\mu} {\approx}$ 1.5) at $\hbar\omega <$ 0.25 MeV and increase slowly at $\hbar\omega >$ 0.25 MeV. The occupation probability of orbital $\pi{5/2}$ ($i_{13/2}$) drops down gradually from one to nearly zero from about 0.25 MeV to 0.30 MeV. The occupation of proton orbitals $\pi{1/2}$ ($f_{7/2}$), $\pi{3/2}$ ($i_{13/2}$) and $\pi{5/2}$ ($i_{13/2}$) leads to the upbending of proton alignment at $\hbar\omega \approx$ 0.25 MeV for the parity-doublet bands in \element{237}U [see Fig. \ref{Fig6}(c)]. \par

For \element{239}Pu, the proton occupation probability of high-$j$ intruder orbital $\pi{5/2}$ ($i_{13/2}$) increases slightly at $\hbar\omega <$ 0.20 MeV and decreases slightly at $\hbar\omega >$ 0.20 MeV. The occupation probability of orbital $\pi{5/2}$ ($f_{7/2}$) decreases slowly at $\hbar\omega <$ 0.20 MeV and slow increase at $\hbar\omega >$ 0.20 MeV. These lead to the angular momentum alignments keep nearly constant at $\hbar\omega <$ 0.20 MeV and slowly decrease at $\hbar\omega >$ 0.20 MeV in the parity-doublet bands in \element{239}Pu [see Figs. \ref{MoI and Jx}(g) and \ref{MoI and Jx}(h)].

In order to clearly understand the upbending mechanism, the contributions of protons to the angular momentum alignments $\langle J_x \rangle$ in the parity-doublet bands in \element{237}U and \element{239}Pu are shown in Fig. \ref{Fig6}. Fig. \ref{Fig6}(a) shows that the sharp increase for the parity-doublet bands in \element{237}U mainly comes from the off-diagonal part of proton alignments. However, as shown in Fig. \ref{Fig6}(b) that both the diagonal and off-diagonal parts vary smoothly for the rotational bands in \element{239}Pu.

The contribution of interference terms $j_x(\mu\nu)$ between the proton orbitals $\mu$ and $\nu$ to the off-diagonal angular momentum alignments in the parity-doublet bands in \element{237}U and \element{239}Pu are shown in Figs. \ref{Fig6}(c) and \ref{Fig6}(d). The interference terms  $j_x(\mu\nu)$ between orbitals of proton $\pi^{2}i_{13/2}f_{7/2}$ pairs are denoted by black solid lines, other interference terms are denoted by black dotted lines. The total interference term from all the proton $\pi^{2}i_{13/2}f_{7/2}$ pairs are denoted by olive dashed lines.

It can be seen clearly that the upbending at $\hbar\omega \approx$ 0.25 MeV in the parity-doublet bands  in \element{237}U are mainly due to the suddenly gained alignment of protons occupying orbital $\pi {1/2}$ $(f_{7/2})$ and high-$j$ intruder orbitals $\pi {3/2}$ $(i_{13/2})$ and  $\pi {5/2}$ $(i_{13/2})$. As shown in Fig. \ref{Fig6}(c), the interference terms between $\pi^{2}i_{13/2}f_{7/2}$ pairs play a very important role in the sharp increased alignment. For \element{239}Pu, there is little contribution from the interference terms $j_x(\mu\nu)$ from $\pi^{2}i_{13/2}f_{7/2}$ pairs at $\hbar\omega <$ 0.20MeV, which decreases slightly at $\hbar\omega >$ 0.20MeV. Therefore, the angular momentum alignments of the parity-doublet rotational bands in \element{239}Pu are quite plain. According to the discussion above, the interference terms of proton octupole correlation pairs of $\pi^{2}i_{13/2}f_{7/2}$ play a key role in the rotational properties in the rotational bands between \element{237}U and \element{239}Pu.

\subsection{Simplex splittings in parity-doublet bands of \element{237}U and \element{239}Pu}

From Figs. \ref{MoI and Jx}(a) and \ref{MoI and Jx}(b), one can see that the kinematic moments of inertia and their variations with rotational frequency are very different between $s=+i$ and $s=-i$ bands for \element{237}U. There is a simplex splitting occurred at low rotational frequency ($\hbar\omega <$ 0.10 MeV) region.  The similar phenomenon is occurred in the parity-doublet rotational bands in \element{239}Pu in Figs. \ref{MoI and Jx}(c) and \ref{MoI and Jx}(d). These simplex splittings in the parity-doublet rotational bands of \element{237}U and \element{239}Pu can be understood from the behavior of the unpaired neutron $\nu{1/2}$ ($d_{5/2}$) orbital.

\begin{figure}[!htbp]
	\includegraphics[scale=0.33]{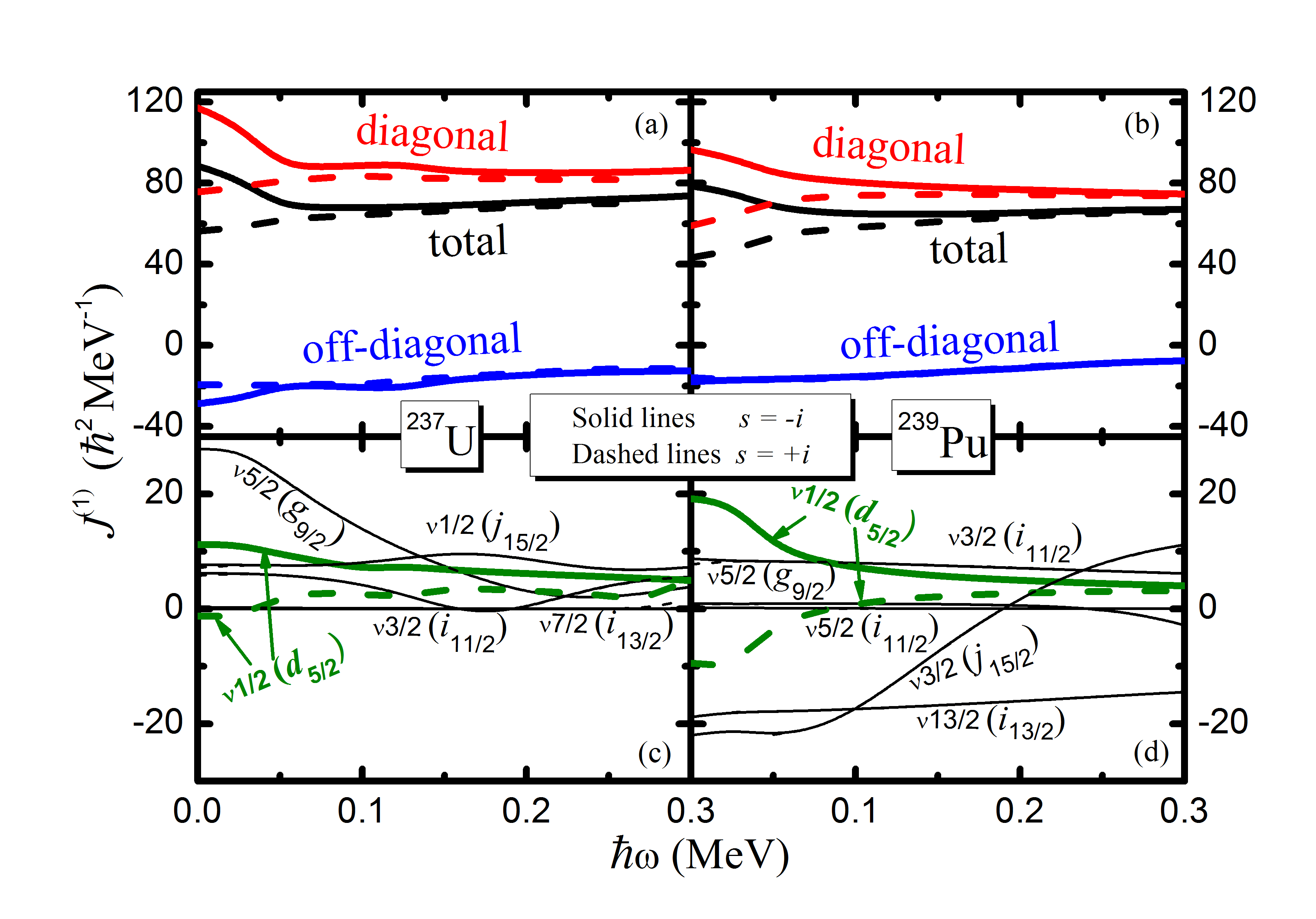}
	\caption{(Color online) Contributions of neutrons to the moments of inertia $J^{(1)}$ for the parity-doublet bands in \element{237}U and \element{239}Pu. The diagonal $\sum\nolimits_{\mu}{j_x}({\mu})$ and off-diagonal parts $\sum\nolimits_{{\mu}<{\nu}}{j_x}({\mu}{\nu})$ are denoted by red and blue lines, respectively (top row). The contribution of nucleon occupying each neutron orbitals are presented in bottom row.}
	\label{Contribution of single particle for Jx}
\end{figure}

As it shows in Figs. \ref{Neutron_Occupation}(a)-(d) that the neutron occupation probability in simplex $s=+i$ and $s=-i$ bands for both \element{237}U and \element{239}Pu are very similar. The neutron $\nu{1/2}$ orbital is half occupied ($n_{\mu} \approx$ 1), and other neutron orbitals are nearly full occupied ($n_{\mu} \approx$ 2) or empty ($n_{\mu} \approx$ 0). The contributions of neutrons to the moments of inertia $J^{(1)}$ in the parity-doublet bands in \element{237}U and \element{239}Pu are shown in Fig. \ref{Contribution of single particle for Jx}, in which the contributions from the diagonal and off-diagonal parts in the $s=+i$ and $s=-i$ bands are presented in Figs. \ref{Contribution of single particle for Jx}(a) and \ref{Contribution of single particle for Jx}(b), respectively. It can be seen clearly that the simplex splittings in the moments of inertia mainly come from the contributions of the diagonal part.

The diagonal parts $J^{(1)}$  of contribution from each neutron orbitals near the Fermi surface for the parity-doublet rotational bands in \element{237}U and \element{239}Pu are presented in Figs. \ref{Contribution of single particle for Jx}(c) and \ref{Contribution of single particle for Jx}(d). It can be seen that the contribution of each simplex doublet is almost same to each other except for the $\nu{1/2}$ $(d_{5/2})$ orbitals, which shows significant splittings at $\hbar\omega <$ 0.10 MeV. It is clearly shows that the simplex splittings of $J^{(1)}$ at $\hbar\omega <$ 0.10 MeV for the parity-doublet bands in $^{237}$U and $^{239}$Pu come from the splitting of the contribution of $\nu{1/2}$ $(d_{5/2})$ orbitals.

\section{Summary}{\label{Sec:summary}}

The parity-doublet rotational bands in odd-$A$ isotones \element{237}U and \element{239}Pu are investigated by using the cranked shell model with the pairing correlation treated by a particle-number-conserving method, in which octupole deformation are taken into account in reflection-asymmetric nuclear system. The experimental moments of inertia and alignments versus the rotational frequency $\hbar\omega$ are reproduced very well by the PNC-CSM calculations. The microscopic mechanism of the distinct difference of rotational behaviors between the bands in \element{237}U and \element{239}Pu, and simplex splittings in the parity-doublet bands for both of \element{237}U and \element{239}Pu are explained.

Sharp upbendings of the moments of inertia $J^{(1)}$ occur at $\hbar\omega \approx$ 0.25 MeV in the parity-doublet bands in \element{237}U whereas $J^{(1)}$ is quite plain during the whole observed frequency for \element{239}Pu. This is due to the contribution of the alignments from nucleon occupying the octupole-correlation pairs of $\pi ^{2} i_{13/2}f_{7/2}$. The upbendings of the parity-doublet bands in \element{237}U are due to the alignments of the interference terms of protons occupying the $\pi f_{7/2}$ (${1/2}$) and the high-$j$ intruder $\pi i_{13/2}$ $(1/2, 3/2)$ orbitals while it contributes a little to the parity-doublet bands in \element{239}Pu.

There are splittings of the simplex partner bands at $\hbar\omega <$ 0.10 MeV of the parity-doublet bands in both \element{237}U and \element{239}Pu. It is found that the simplex splittings between the $s=+i$ and $s=-i$ rotational bands result from the contributions of alignment of neutron occupying the $\nu{1/2}$ ($d_{5/2}$) orbital.

\begin{acknowledgements}
This work is supported by the National Natural Science Foundation of China (Grant Nos. U2032138, 11775112).
\end{acknowledgements}

\bibliographystyle{apsrev4-1}
\bibliography {references}

\end{document}